\def   \ni {\noindent}

\def   \ssk {\vskip  5truept}

\def   \bsk {\vskip 15truept}

\def   \newline {\hfil\break}

\documentstyle[epsfig]{article}
\begin{document}

\hsize 5truein
\vsize 8truein
\font\abstract=cmr8
\font\keywords=cmr8
\font\caption=cmr8
\font\references=cmr8
\font\text=cmr10
\font\affiliation=cmssi10
\font\author=cmss10
\font\mc=cmss8
\font\title=cmssbx10 scaled\magstep2
\font\alcit=cmti7 scaled\magstephalf
\font\alcin=cmr6 
\font\ita=cmti8
\font\mma=cmr8
\def\ref{\par\noindent\hangindent 15pt}
\null


\title{\ni COSMIC GAMMA-RAY BURSTS: THE REMAINING MYSTERIES
}                                               

\bsk \bsk
\author{\ni K.~Hurley $^{1}$}                                                       
\bsk
\affiliation{1) UC Berkeley, Space Sciences Laboratory, Berkeley CA 94720-7450, USA
}                                                
\bsk
\baselineskip = 12pt

\abstract{ABSTRACT \ni
To anyone who has read a scientific journal or even a newspaper in the
last six months, it might appear that cosmic gamma-ray bursts hold no more
mysteries: they are cosmological, and possibly the most powerful explosions
in the Universe.  In fact, however, bursts remain mysterious in many ways.
There is no general agreement upon the nature of the event which releases
the initial energy.  One burst at least appears to strain the energy budget
of the merging neutron star model.  There is evidence that another recent event
may have come from a nearby supernova.  Finally, while the number count statistics
clearly show a strong deviation from the -3/2 power law expected for a Euclidean,
homogeneous distribution, the distributions of some classes of bursts appear to follow 
a -3/2 power law rather closely.  The recent data on bursts is reviewed, some of the mysteries discussed, and future experiments are outlined.  
}                                                    
\bsk
\baselineskip = 12pt

\bsk
\baselineskip = 12pt


\text{\ni 1. THE STORY SO FAR
\ssk
\ni     
Until about one year ago, the gamma-ray burst (GRB) distance scale was completely
unknown, and, for most practical purposes, unconstrained.  On February 28, 1997,
the BeppoSAX spacecraft detected and rapidly localized a burst with the WFC, and,
pointing the NFI instruments at the location, discovered its first fading X-ray 
counterpart (figure 1: Costa et al. 1997).  Shortly thereafter, van Paradijs (1997) and his 
colleagues observed the region with an optical telescope, and discovered the first
fading optical counterpart associated with a GRB.  The X-ray and optical
positions were confirmed by the Interplanetary Network (Hurley et al. 1997), and
a twenty year search for GRB counterparts had come to an end.

\begin{figure}
\centerline{\epsfig{file=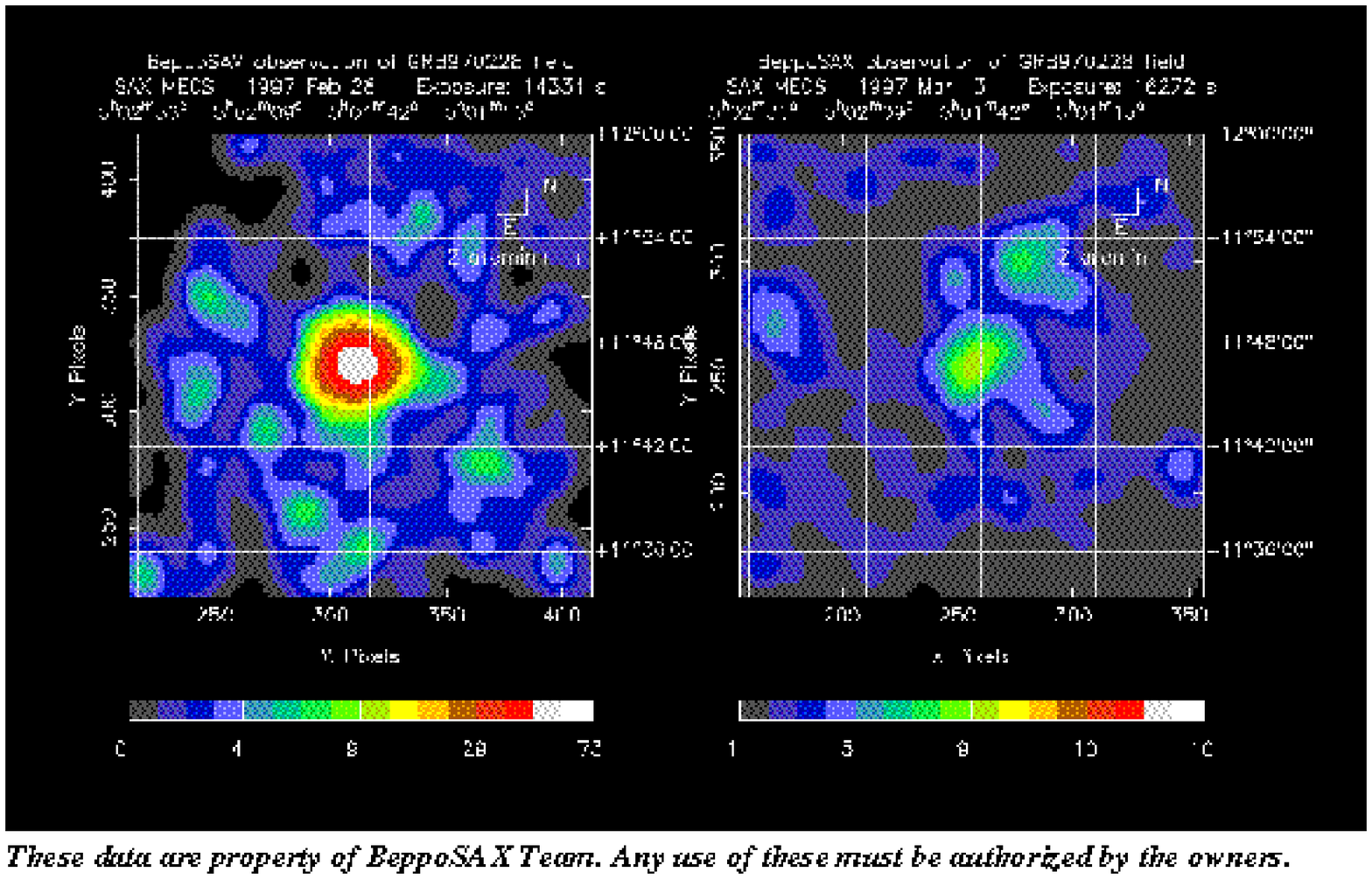, width=7cm}}
\caption{FIGURE 1. BeppoSAX NFI images of the fading 2-10 keV X-ray source associated with GRB 970228, 8 and
48 hours after the burst, from Costa et al. 1997.  
}
\end{figure}


Today we know that most, if not all GRB's are accompanied by fading X-ray counterparts,
and that approximately 50\% of them also have fading optical counterparts.  The fading
generally can be described by a power law in time with an index $\approx$-1.5, although
there are noteworthy exceptions (different power law indices, and fading behavior
which is not a monotonic decrease).

The redshift of GRB970228 is still unknown.  The first GRB redshift was measured, or
more accurately, constrained, by Metzger (1997).  From the observation of an absorption
line system towards the source of GRB970508, he was able to show that z$>$0.8.  From the
lack of a Ly $\alpha$ forest, z$<$2.2.  This remains one of only three redshifts measured
to date.  The other two are z=0.96 (Djorgovski et al. 1998) and z=3.4 (Kulkarni et al. 1998a).

The behavior of GRB afterglows can be understood in terms of the cosmological
fireball model, in which a relatistic blast wave moving with a bulk Lorentz
factor $\Gamma > 100$ impacts matter external to the burst source, resulting
in the shock acceleration of particles and synchrotron emission (e.g. Wijers et al.
1997)
Indeed, radio observations of GRB970508 by Frail et al. (1997) provide support for
this picture.  Figure 2 shows a VLA observation of the radio emission from the
source as a function of time.  The emission shows evidence for ``scintillation'',
which gradually disappears.  The explanation (Goodman 1997) is analogous to the explanation
for the fact that stars twinkle, while planets do not.  In the GRB case the scintillation
is caused by scattering in our galaxy's interstellar medium, and it stops when the
fireball has expanded to $\sim3\mu$arcseconds.  At this point, the expansion
velocity has slowed to $\Gamma\sim 1$.

\begin{figure}
\centerline{\epsfig{file=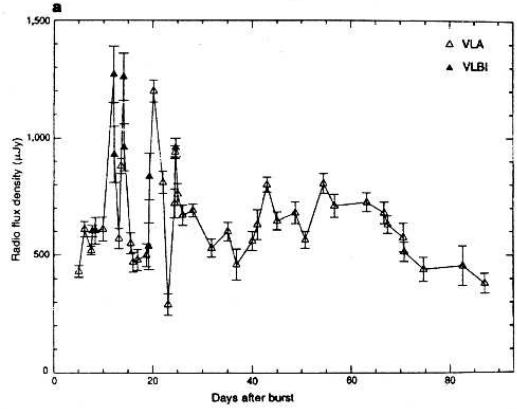, width=7cm}}
\caption{FIGURE 2. Flux of the fading radio counterpart to GRB970508 (Frail 1997).  
Note the presence of scintillations which damp out, followed by a smoother, monotonic
decay.    
}
\end{figure}

\bsk
\ni 2. NEUTRON STAR MERGERS AND HYPERNOVAE
\ssk
\ni 

One possibility for generating the GRB energy involves the merging of a neutron
star in a binary system.  The binary companion is either another neutron star
or a black hole (Paczynski 1991).  If the resulting gamma-radiation is not beamed, 
this need only occur once every 10$^6$ years in every galaxy to explain the
number of bursts observed.  Such a merger will liberate 10$^{52}$ erg of energy.
Considering that the luminosity of the entire Universe is just 10$^{53}$ erg/s, 
mostly in the optical band, this is
indeed a phenomenal energy.  The burst rate should roughly follow that of star formation,
which is thought to peak around z$\approx$1.25 (Madau et al. 1996).  The timing would be approximately
the following.  Once two massive stars have formed, the first might go supernova
in $\approx$ 10$^7$ y, leaving behind a high mass X-ray binary system.  The lifetime
of this system would be short, only $\approx$ 10$^5$ y, before the second star explodes
as a supernova, leaving a binary neutron star system.  This system will radiate its
orbital energy away as gravitational radiation in several times 10$^9$ y, resulting
in a merger.

If this idea is basically correct, we might expect that many mergers will be
associated with the
small, low mass (and therefore intrinsically dim) galaxies which have undergone short
bursts of star formation.  In our own galaxy, there is some evidence that the binary
pulsars have high velocities, $>$200 km/s.  In the low mass galaxies where GRBs may
be generated, a binary neutron star system can therefore achieve escape velocity,
and in only $\approx 10^8$ y, it can reach a distance $\approx 30$ kpc from its
host galaxy.  Thus by the time the merger occurs, the system need not be in a galaxy
at all.

There is, however, another possibility for the energy release.  There is a
strict upper limit to the amount of energy which can be released by two merging
neutron stars: E=mc$^2\approx5x10^{53} erg$, and of course not all of this
can go into the radiation which we observe.  Some bursts, like GRB971214 (z=3.4:
Kulkarni et al. 1998a) strain this energy budget if they are emitting isotropically.
They may be caused by ``hypernovae'', extremely energetic supernovae which result
from the collapse of a 10 - 15 M$_{\odot}$ star to a black hole with an
accretion disk which produces the energy for the burst.

Unlike the binary neutron stars, the massive stars which are the progenitors
of hypernovae live only $\approx 10^6$ y, and do not travel far from their
birthplaces.  This leads to an observational test which can distinguish the
two possibilities.  Bursts generated by merging neutron stars should occur
outside their host galaxies fairly often, while bursts generated by hypernovae
should appear within them.  So far, there is some observational evidence that
favors the hypernova model, but this is based on only some 8 bursts, and more
data will be required for a decisive test.  Indeed, in these 8 cases, the host
galaxies have not always been directly detected.  In some cases, the presence
of a host galaxy has been inferred from the flattening of the optical counterpart
light curve (e.g. Groot et al. 1998), which is interpreted as due to the presence
of an underlying galaxy which has too low a surface brightness to detect directly. 

The approximate energy budget for a GRB, in either case, is given in Table 1.
Starting from, say, 10$^{52}$ erg, this table shows the amount of energy which
goes into various emissions.  Neutrinos are expected to carry away much of
the intrinsic energy, but there are no measurements at present which constrain
them significantly.  Similarly, the optical and radio emissions during the burst
could be significant, but they have not been measured.

\begin{table}[h]
\begin{center}
\caption{Table 1. The Approximate GRB Energy Budget}
\end{center}
\vspace{0.4 in}
\begin{tabular}{cc}
DURING THE BURST & AFTERGLOW \\
$\rm \nu:  >10\%(?)$ & X-rays  10\% \\
$\gamma$ rays: 70\% & Optical: 2\% \\
X-rays: 8\% & Radio: 0.05\% \\
Optical: ? &  \\
Radio: ? &  \\
\end{tabular}
\end{table}

\bsk
\ni
3. Mysteries
\ssk
\ni

Even though the GRB distance scale appears to be resolved, the source of
the energy, as discussed above, remains a matter of some speculation.  There
are numerous other unknowns, too, three of which are now discussed.

\bsk
\ni
3.1 Presence and absence of counterparts
\ssk
\ni

Although virtually all GRB's seem to have fading X-ray counterparts, less
than 60\% display fading optical counterparts, and perhaps only 33\% display
radio counterparts.  Furthermore the presence or absence of an optical counterpart
is not correlated with the GRB intensity or the intensity of the fading counterpart.
The reason for this may be related to extinction (Reichart 1998).  X-rays are attenuated roughly
as $\nu^{-3}$, where $\nu$ is the frequency, while optical radiation is attenuated
as $\sim \nu^1$.  If the spectrum of a counterpart is redshifted, the ratio of
optical to X-ray optical depths will go approximately as $(1+z)^4$, where z is the redshift.
Thus the optical light may be considerably attenuated with respect
to the X-rays, even at modest redshifts.  This is one more piece of circumstantial
evidence that GRBs may originate close to the birthplace of their progenitors, in
dusty, star-forming regions, and therefore be related to hypernovae rather than
binary neutron stars.

\bsk
\ni
3.2 A GRB-Supernova connection?
\ssk
\ni

The search for the optical counterpart to GRB980425 revealed that the
8' radius BeppoSAX WFC error circle (Soffitta et al. 1998) contained
an optical transient source which was in fact a supernova (figure 3: Galama et al.
1998).  The supernova, 1998bw in the galaxy ESO 184-G82, is an unusual
one (Filippenko 1998; Kulkarni et al. 1998b), and the galaxy is close, with a redshift
of z=0.008 (Kay et al. 1998).

\begin{figure}
\centerline{\epsfig{file=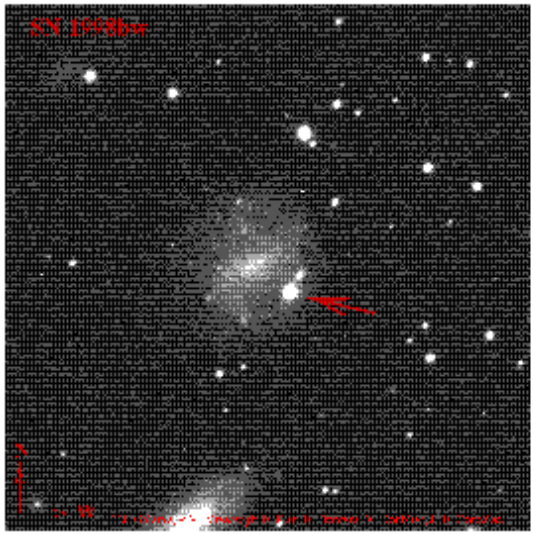, width=7cm}}
\caption{FIGURE 3. SN1998bw in ESO 184-G82, which may be associated with
GRB9890425.    
}
\end{figure}

Pian et al. (1998) found two X-ray sources in the WFC error circle using
the BeppoSAX NFI.  One is variable, and the other is constant, but only the constant source 
has a position consistent with the supernova (Piro et al. 1998).

On the one hand, it would be tempting to believe that this is a hypernova
and that the GRB-SN association is indeed true.  On the other hand, though,
bursts from supernovae were among the first theories proposed to explain
GRB (Colgate 1974) and for twenty years, temporal and spatial coincidences
between the two were searched for, with no positive results.  This association
is still debatable (Wang and Wheeler 1998; Kippen et al. 1998).

\bsk
\ni
3.3 Euclidean and non-Euclidean GRBs
\ssk
\ni

It is now well accepted that the log N-log S, or number count distribution of
gamma-ray bursts \it as a whole \rm deviates strongly from the -3/2 power law that one would
expect if the sources were distributed homogeneously in Euclidean space.  The
fall-off is attributed to redshift effects.  However, a number of authors
have pointed out that certain categories of bursts, defined by their energy
spectra and durations, for example, display nearly Euclidean distributions 
(Pizzichini 1995; Belli 1997; Tavani 1998).  An example is shown in figure 4:
the long duration, soft-spectra bursts and the short duration bursts follow
a -3/2 power law more closely than the ensemble of all bursts.  This is
counterintuitive, since, in a cosmological distribution, bursts with soft spectra
would be expected to be more distant and display a strong deviation from
a Euclidean distribution.

\begin{figure}
\centerline{\epsfig{file=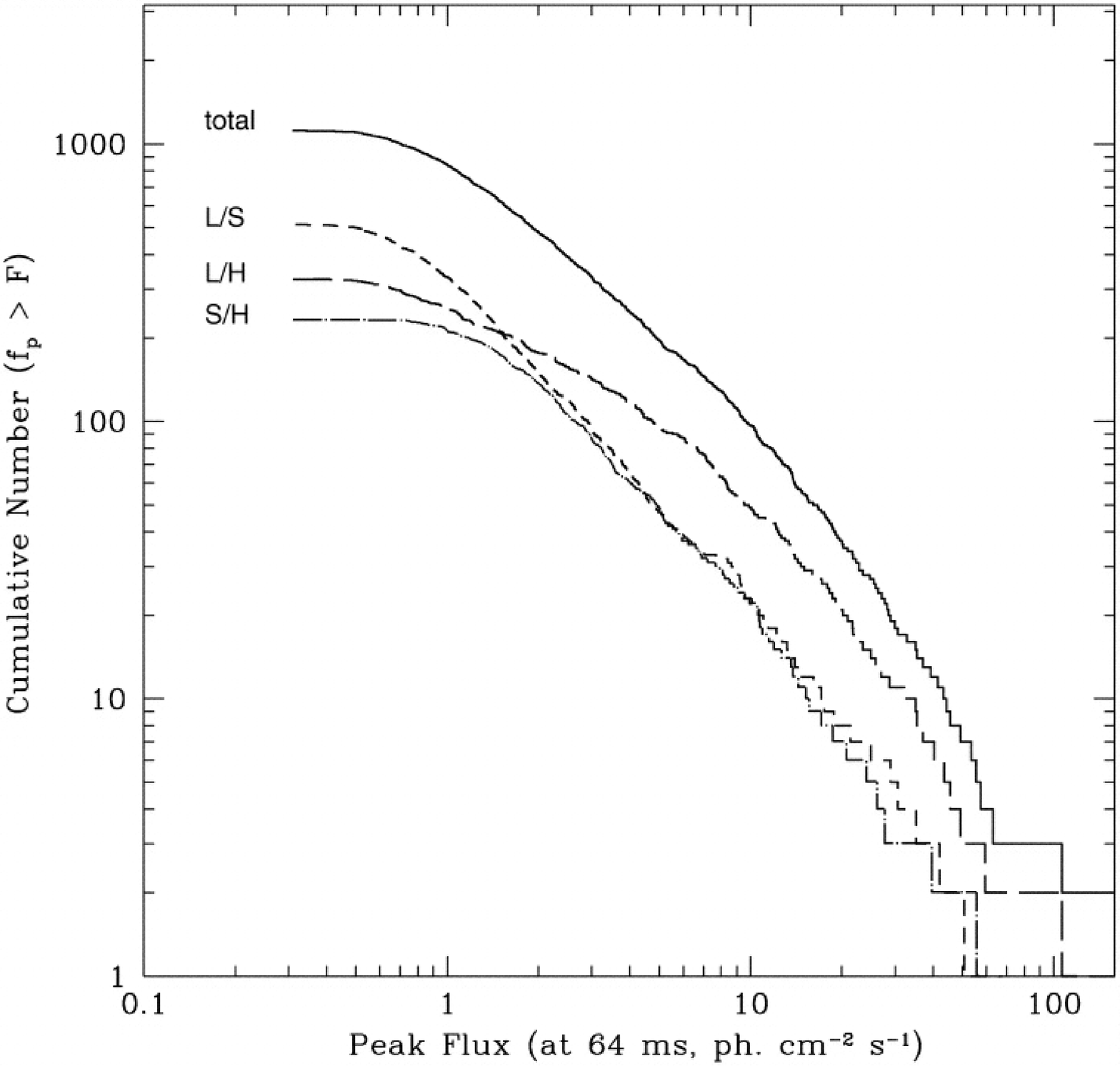, width=7cm}}
\caption{FIGURE 4. The number-intensity relations for various categories of
bursts, from Tavani (1998).  L/S, L/H, and S/H refer to long/soft, long/hard,
and short/hard, respectively.    
}
\end{figure}

There are many possible solutions: selection effects, multiple populations, and/or
strong luminosity and spectral evolution.  It should also be noted that no
short duration bursts have yet been associated with optical counterparts, so
their distance scale could be quite different.  In fact, they could even
be galactic since, although their spatial distribution is isotropic, the 
agreement between their log N-log S distribution and a Euclidean distribution
indicates that the population has not been fully sampled.

\bsk
\ni 4. CONCLUSIONS 
\ssk
\ni 

From this brief review, a number of important questions can be identified.\\

1. Are GRBs in their host galaxies, or outside them?\\

2. What is the distribution of their distances?\\

3. What is the intrinsic luminosity function for bursts?\\

4. Are there different classes of bursts, as the data seem to suggest?\\

5. What is the multiwavelength behavior of GRB light curves immediately
after the burst?\\

Answering these questions will require hundreds of rapidly determined, small
GRB error boxes, followed up by multiwavelength observations.  Elsewhere in
these proceedings (Hurley 1998), the missions capable of providing these
error boxes have been summarized.  In the immediate future, we have BeppoSAX,
BATSE, and the 3rd Interplanetary Network.  Around 1999, HETE-II should be
launched; its nominal lifetime is 2 years.  By 2003, a MIDEX mission dedicated
to GRB studies may be in operation.  INTEGRAL, in the years 2001-2003, will
fill a void.  Not only will it have an inherent GRB capability from the IBIS
experiment, but also it can act as a point in a future interplanetary network.
As such, it will serve the important role of maintaining interest and expertise
in this important and exciting field.


{\references \ni REFERENCES
\ssk

\ref Belli, B. 1997, Ap. J. 479, L31
\ref Colgate, S. 1974, Ap. J. 187, 333
\ref Costa, E., et al. 1997, Nature 387, 783
\ref Djorgovski, S. et al. 1998, GCN GRB Observation Report 137
\ref Filippenko, A. 1998, IAUC 6969
\ref Frail, D. et al. 1997, Nature 389, 261
\ref Galama, T. et al. 1998, Nature 395, 670
\ref Goodman, J. 1997, New Astronomy 2, 449
\ref Groot, P. et al. 1998,  Ap. J. 502, L123
\ref Hurley, K., et al. 1997, Ap. J. 485, L1
\ref Hurley, K. 1998, these proceedings
\ref Kay, L. et al. 1998, IAUC 6969
\ref Kippen, R. et al. 1998, Ap. J. 506, L27
\ref Kulkarni, S. et al. 1998, Nature 393, 35
\ref Kulkarni, S. et al. 1998, Nature 395, 663
\ref Madau, P. et al. 1996, MNRAS 283, 1388
\ref Metzger, M., et al. 1997, Nature 386, 686
\ref Paczynski, B. 1991, Acta Astronomica 41, 257 
\ref Pian, E. et al. 1998, GCN GRB Observation Report 61
\ref Piro, L. et al. 1998, GCN GRB Observation Report 155
\ref Pizzichini, G. 1995, Proc. 24th ICRC (Rome), OG2.1.8, p. 81
\ref Reichart, D. 1998, Ap. J. 495, L99
\ref Soffitta, P. et al. 1998, IAUC 6884
\ref Tavani, M. 1998, Ap. J. 497, L21
\ref van Paradijs, J. et al. 1997, Nature 386, 686
\ref Wang, L., and Wheeler, C. 1998, Ap. J. 504, L87
\ref Wijers, R., Rees, M., and Meszaros, P. 1997, MNRAS 288, L51
}                      

\end{document}